\documentclass[12pt]{article}
\usepackage[cm]{fullpage}
\usepackage{amsmath}
\usepackage{amsfonts}
\usepackage{amssymb}
\usepackage{amsthm}
\usepackage{graphicx}
\usepackage{fancyhdr}
\usepackage{xcolor}
\usepackage{bbding}
\usepackage{multicol}

\definecolor{bb}{rgb}{0.3, 0.5, 1}
\definecolor{bg}{rgb}{0.1, 0.1, 0.5}
\usepackage[%
    pdftitle={The title of your letter},
    pdfauthor={Your name},
    pdfsubject={The subject of the letter},
    pdfkeywords={Any keywords},
     colorlinks=true,  
     linkcolor=bg, 
     citecolor=bg, 
     urlcolor=bg, 
     pdfnewwindow=true,
     pagebackref=true,
     filecolor=bb,
     pdffitwindow=false,     
     pdfstartview={FitH}    
]{hyperref}

\def\ba{\begin{eqnarray}}
\def\ea{\end{eqnarray}}
\def\be{\begin{equation}}
\def\ee{\end{equation}}

\def\L{\mathcal{L}}
\def\K{\mathcal{K}}

\def\nn{\nonumber}

\def\d{\mathrm{d}}

\def\mn{_{\mu \nu}}

\def\l{\ell}

\def\({\left(}
\def\){\right)}
\def\ie{{\it i.e. }}

\def\nn{\nonumber}
\def\p{\partial}

\def\mf{M_5^3{}}
\def\mq{M_4^2{}}

\def\pp{\partial_\mu\partial_\nu \pi}
\def\dbi{1+(\p \pi)^2}
\def\tg{\tilde\gamma}

\begin{document}


\begin{flushright}
\end{flushright}
\vskip 0.9cm

\centerline{\Large \bf DBI and the Galileon reunited}
\vskip 0.7cm
\centerline{\large Claudia de Rham$^\clubsuit$
 and Andrew J. Tolley$^\spadesuit$}
\vskip 0.3cm

\centerline{\em $^\clubsuit$\,D\'epartment de Physique  Th\'eorique, Universit\'e
de  Gen\`eve,}
\centerline{\em 24 Quai E. Ansermet, CH-1211,  Gen\`eve, Switzerland}

\centerline{\em $^\spadesuit$\, Perimeter Institute for Theoretical Physics, Waterloo ON}
\centerline{\em 31 Caroline street North, N2L 2Y5, Canada.}

\vskip 1.9cm

\begin{abstract}
We derive the relativistic generalization of the Galileon, by studying the brane position modulus
of a relativistic probe brane embedded in a five-dimensional bulk. In the appropriate Galilean contraction limit, we recover the complete Galileon generalization of the DGP decoupling theory and its conformal extension. All higher order interactions for the Galileon and its relativistic generalization naturally follow from the brane tension, induced curvature, and the Gibbons-Hawking-York boundary terms associated with all bulk Lovelock invariants.
 Our approach makes the coupling to gravity straightforward, in particular allowing a simple rederivation of the nonminimal couplings required by the Covariant Galileon.
The connection with the Lovelock invariants makes the well-defined Cauchy problem manifest, and gives a natural unification of four dimensional effective field theories of the DBI type and the Galileon type.
\end{abstract}

\vspace{1cm}


\newpage

\tableofcontents

\newpage

\section{Introduction}

Infrared modified theories of gravity such as the Dvali-Gabadadze-Porrati (DGP) \cite{DGP} model and its generalizations \cite{gigashif,cascading,aux1,intersecting,Kaloper:2007ap}, may at first sight have little to do with stringy motivated models of inflation such as Dirac-Born-Infeld (DBI) inflation, \cite{Silverstein:2003hf}. However, from the perspective of their four dimensional effective theories they belong to the same class of models, theories of scalar fields for which certain types of derivative interactions are allowed to be {\it large}, at least for classical field configurations \cite{nima,lpr}. In a typical effective field theory, the derivative interactions, being irrelevant operators, cannot be taken to be large without going outside the regime of validity of the effective field theory. What makes these theories different is the existence of a symmetry, albeit one that is nonlinearly realized, that protects the form of the interactions. Both theories have in common that there are only a finite number of allowed interactions, which are invariant under the required symmetries, and lead to second order equations of motion. The precise way in which these theories are believed to make sense beyond their na\"ive regime of validity is different in each case. In particular, infrared (IR) modified gravities rely on the idea of the Vainshtein/kinetic chameleon effect to raise the effective cutoff scale in the presence of a source, in turn suppressing quantum corrections.

In the case of DBI, the symmetry is the nonlinearly realized higher dimensional Poincar\'e or Anti-de Sitter (AdS) invariance, \cite{Aharony:1999ti}. This symmetry is clearly inherited from the higher dimensional picture of the scalar describing the modulus associated with a D-brane moving in a warped throat. By contrast, in the DGP case, the symmetry is the Galilean limit of the Poincar\'e or AdS symmetry. The symmetry is similarly inherited from the extra dimensional Poincar\'e symmetry, but is not identical because of the very different nature of the DGP decoupling limit. The most general theory which exhibits this Galileon symmetry, and gives equations of motion with a well-defined Cauchy problem, was recently derived in \cite{Nicolis:2008in}, while its coupling to gravity was presented in \cite{Deffayet:2009wt}. In this work, the allowed interactions for the Galileon were determined by demanding that the equations of motion remain second order in derivatives,
and an elegant motivation was then provided in \cite{Deffayet:2009mn}, using the anti-symmetric Levi-Civita symbol.
The cosmological phenomenology was considered in \cite{Chow:2009fm} and many other aspects of field theories of this type have been considered in \cite{generic}. In particular the absence of stable solitonic solutions was recently shown in  \cite{Endlich:2010zj}.

In what follows, we will complete the connection, both by demonstrating that there is a unified class of 4 dimensional effective theories, having the same symmetry as the DBI action, which in the appropriate limit reproduces all the interaction terms of the Galileon, and by giving a clear prescription for finding a higher dimensional theory for which these effective theories are the appropriate decoupling limit. We also emphasize that this approach makes the coupling to gravity as well as the multifield extension  straightforward.

The method of deriving the Galileon effective action by considering a probe brane in 5 dimensions was already suggested in \cite{Nicolis:2008in}. In particular in their section 3.1, this ideology was utilized to try to obtain the Galileon interactions from functions of the intrinsic curvature of the metric $g_{\mu\nu}=e^{2\pi}\eta_{\mu\nu}$ which automatically preserves the conformal group $SO(2,4)$, giving rise to the conformal Galileon. However, our approach will be different, since we work with the true induced metric,\footnote{The different exponent in the exponential is just a matter of convention.} $g_{\mu\nu}=e^{-2\pi/\l}\eta_{\mu\nu}+\partial_{\mu}\pi \partial_{\nu} \pi$ for which the conformal group symmetry $SO(2,4)$ is nonlinearly realized in a different manner than that considered in \cite{Nicolis:2008in}, and consider also functions of the extrinsic curvature that arise as the Gibbons-Hawking-York boundary terms of the Lovelock invariants which are generically generated in the bulk. What we gain from this is a straightforward way to realize this in extra dimensions, and a manifest way to couple to gravity without introducing higher derivative terms, since we only utilize the famous Lovelock invariants in 4 and 5 dimensions. This means that we do not need to take the Galilean limit to see that the Cauchy problem is well-defined. We also find that all the allowed interactions can be obtained by utilizing a remarkable recursion relation, whose origin is in the 5 dimensional Gauss-Codazzi equations.

The simplicity of our result is encoded in the fact that the Lagrangian for the probe brane that is sufficient to reproduce all the interactions of the Galileon, Conformal Galileon, their relativistic generalizations, and also to account for the coupling to gravity is given by the remarkably simple expression:
\ba
\nonumber
\hspace{-5pt}{\cal L}= \sqrt{-g}\(-\lambda+\frac{\mq}{2} R-\mf K-\beta \frac{\mf}{m^2} \mathcal{K}_{\rm GB}\)+\mathcal{A}(\pi)\,,\hspace{10pt}
\ea
where $\mathcal{A}(\pi)$ is a `tadpole' contribution whose origin may be in integration over the bulk action or, as for DBI, Chern-Simons terms in the effective action, and
the term
$\mathcal{K}_{\rm GB}$
is simply the Gibbons-Hawking-York boundary term associated with a bulk Gauss-Bonnet term. Here $g_{\mu\nu}$ and $K_{\mu\nu}$ are the induced metric and extrinsic curvature of the brane, and the curvature invariants are functions of the induced metric and $\pi$ is the position modulus of the brane. The presence of terms which are odd in the extrinsic curvature in the probe brane action appears peculiar. For an orbifold brane terms of this form are familiar, since we may work with the theory defined only on the right hand side of the brane for example, and these terms are seen to arise as Gibbons-Hawking-York boundary terms of the bulk contributions. Alternatively in the double cover picture, the fact that the first derivative of the bulk metric is discontinuous at the position of the brane means that boundary terms on each side do not cancel.
For a dynamical probe brane moving in a smooth geometry however, these terms are usually canceled by the boundary terms on the other side of the brane. An exception is when the brane represents the boundary between two asymmetric regions. 

Since our results are derived in a probe brane approximation, we will not be able to infer the usual DGP decoupling limit in the conventional sense. In the DGP decoupling limit, a kinetic term for $\pi$ is only obtained from a mixing with gravity, whereas  in the following, the kinetic term for $\pi$ will arise principally from the tension (although in the AdS case additional contributions will come from the higher order invariants). Furthermore, $\pi$ has a different meaning, in the probe brane limit $\pi$ is the position modulus of the brane. In DGP this degree of freedom is projected our because of the orbifold symmetry. The degree of freedom $\pi$ represents a type of brane bending mode which really arises from the bulk metric. It is at present an open question whether the full Galileon interactions can arise in some extension of the DGP model via a similar type of decoupling limit.

We proceed as follows: In Section \ref{sec:symmetries} we describe the nonlinear symmetry realized in both the DBI and Galileon models. We then rely on this symmetry to derive the most relevant operators and the generic form of the quantum corrections and present in section \ref{sec:ProbeBrane} the most general induced action on a relativistic probe consistent with these symmetries and having well-defined equations of motion. Moreover, in the appropriate non-relativistic limit we recover the Galileon ansatz. To make contact with the standard warped DBI model, we present in section \ref{sec:AdS} the AdS generalization of the probe brane induced action and emphasize the recursive relation existing between the different terms induced on the brane action ensuring the stability of the theory, before recovering the conformal Galileon limit.
The coupling to gravity is straightforward as shown in section \ref{sec:Covariant}, and is in agreement with results previously obtained in some specific limits. Although gravity necessarily breaks the nonlinear symmetry, we give a prescription for how to replace the symmetry by a larger group using the higher dimensional picture. Finally we review our findings briefly in section \ref{sec:Conclusion}.

\section{Effective Field Theories with nonlinearly realized symmetries}
\label{sec:symmetries}

\subsection{DBI and Poincar\'e invariance}

The DBI probe brane Lagrangian is protected by a nonlinearly realized 5 dimensional Poincar\'e invariance $ISO(1,4)$ (unwarped case) or 5 dimensional AdS invariance $SO(2,4)$ (warped case). Let us first consider the unwarped case for which the Lagrangian takes the form
\be
S=\int \d^4 x\left( -\lambda \sqrt{1+(\partial \pi)^2}+\lambda \right).
\ee
This action is manifestly invariant under the linearly realized $ISO(1,3)$ subgroup ($M_{\mu\nu}$,$P_{\mu}$) of $ISO(1,4)$, \ie the 4 dimensional Poincar\'e group. Furthermore it is invariant under $\pi \rightarrow \pi + c$ which corresponds to translations ($P_4$) in the 5th dimension. The symmetry that is not immediately apparent, is that associated with rotations in the 5th dimension ($M_{4\mu}$). Explicitly the infinitesimal form of the symmetry is
\be
\delta_{v} \pi(x) = v_{\mu} x^{\mu}+ \pi(x) v^{\mu}\partial_{\mu} \pi(x).
\ee
One may check by direct substitution that the Lagrangian shifts by a total derivative under this transformation. However it is more illuminating to consider the transformations of the tensor $
g\mn=\eta\mn+\p_\mu \pi \p_\nu \pi$.
An explicit calculation shows that
\ba
\delta_{v}g\mn&= &\nonumber \partial_{\mu}\pi \partial_{\nu} \delta_{v} \pi +   \partial_{\nu}\pi \partial_{\mu} \delta_{v} \pi   \\
&=& \partial_{\mu}\pi (v_{\nu}+\partial_{\nu}(\pi v^{\alpha}\partial_{\alpha}\pi)) +   \partial_{\nu}\pi (v_{\mu}+\partial_{\mu} (\pi v^{\alpha}\partial_{\alpha}\pi))\nonumber \\
&=& \xi_v^{\alpha} \partial_{\alpha}g\mn+ \partial_{\mu}\xi_v^{\alpha} g_{\alpha \nu}+\partial_{\nu}\xi_v^{\alpha} g_{\alpha \mu},
\ea
where
\be
\xi_v^{\alpha}=v^{\alpha} \pi(x).
\ee
In other words, the transformation on $\pi$ induces a transformation on the metric that looks like a field dependent coordinate transformation. Given this, it is clear that any scalar constructed out of $g\mn$ will transform as a scalar
\be
\delta_{v} S = \xi^{\alpha}_v \partial_{\alpha} S=v^{\alpha} \pi(x)\partial_{\alpha} S.
\ee
Similarly given the definition of the brane extrinsic curvature:
\ba
K\mn=-\frac{\pp}{\sqrt{\dbi}} \, ,
\ea
one may explicitly check that it transforms as a tensor should
\be
\delta_{v}K\mn=\xi^{\alpha}_v \partial_{\alpha}K\mn+ \partial_{\mu}\xi^{\alpha}_v K_{\alpha \nu}+\partial_{\nu}\xi^{\alpha}_v K_{\alpha \mu}.
\ee
In addition, the tadpole term $\pi$ in the Lagrangian is accidently invariant because $\delta_v \pi$ is a total derivative $\delta_v \pi = \partial_{\mu}\( v^{\mu}x^2/2+v^{\mu}\pi^2/2\)$.
Strictly speaking only the equations of motion are invariant, since one cannot in general neglect the boundary term unless it is canceled by appropriate boundary operators. This means there will not in general be a globally conserved charge associated with this symmetry. However, for most of our discussion this subtlety will not matter.
This immediately tells us what the most general form of the Lagrangian that will be invariant under the nonlinearly realized $ISO(1,4)$ symmetry must be of the form
\be
\label{LsymMin}
{\mathcal L} =\sqrt{-g} \, S(g_{\mu\nu},R_{abcd} ,K_{ef}, \nabla_g)+A\, \pi,
\ee
where $S$ is a scalar constructed out of arbitrary functions and covariant derivatives of the Riemann metric associated with $g\mn$, and extrinsic curvature $K\mn$.
There are two ways in which matter can couple to this theory. Matter fields $\chi^i$ which are minimally coupled w.r.t. the metric $g_{\mu\nu}$,
${\cal L}_m(g_{\mu\nu},\chi^i)$
must also transform under the symmetry. For instance for scalars,
\be
\delta_v \chi^i=v^{\mu} \pi \partial_{\mu} \chi^i.
\ee
On the other hand there can be matter fields which are minimally coupled to the metric $\eta_{\mu\nu}$, and consequently will not transform under the symmetry. Of course such fields are decoupled here, but in the covariantized version where gravity is dynamical, they will couple via gravity, and so this distinction will play an important role.

\subsection{DBI and Anti-de Sitter invariance}

The case of most interest in the context of DBI inflation models, is the warped DBI model, where the brane is moving in an approximately Anti-de Sitter throat. Ignoring potential contributions, which mildly break the symmetry, the effective action for the D-brane takes the form
\be
\label{warped}
S=\int \d^4 x\left( -\lambda e^{-4\pi/\l}\sqrt{1+e^{2\pi/\l}(\partial \pi)^2}\pm\lambda e^{-4\pi/\l} \right),
\ee
where we have chosen AdS coordinates in the form
\be
\d s^2=\d y^2+e^{-2y/\l} \eta_{\mu\nu}\d x^{\mu}\d x^{\nu}\,.
\ee
In the inflation literature it is more usual to work with the canonically normalized coordinates $\Phi=\sqrt{\lambda}\,  \l e^{-\pi/\l}$, so that the action (from now on we choose the $+$ sign for the tadpole term corresponding to a supersymmetric D-brane) takes the form
\be
S=\int \d^4 x\left( -\frac{\Phi^4}{\l^4\lambda} \sqrt{1+\frac{\l^4\lambda}{\Phi^{4}}(\partial \Phi)^2}+\frac{\Phi^4}{\l^4\lambda} \right).
\ee
In what follows we shall work with the form given in Eq.~(\ref{warped}). The symmetry in this case is the nonlinearly realized $SO(2,4)$ symmetry, namely the isometry group of $AdS^5$. The generators of $SO(2,4)$, $M_{AB}$ (with $A=-1,0,1,2,3,4$) split up into $M_{\mu\nu}, M_{-1\mu},M_{4\mu},M_{-1,4}$ where $\mu=0,1,2,3$. The $ISO(1,3)$ subgroup is linearly realized via $(M_{\mu\nu},P_{\mu}=(M_{-1\mu}+M_{4\mu})/2)$.
The analogue of the shift symmetry ($M_{-14}$) is the transformation
\be
\delta_c \pi = c\(1 -\frac{1}{\l} x^{\mu}\partial_{\mu} \pi\).
\ee
To see that this is a symmetry, we follow the previous procedure and define the induced metric on the brane as $g_{\mu\nu}=e^{-2\pi/\l} \eta_{\mu\nu}+\partial_{\mu}\pi \partial_{\nu} \pi$.
Then we see that
\ba
\delta_c g_{\mu\nu}=\xi_c^{\alpha} \partial_{\alpha}g\mn+ \partial_{\mu}\xi_c^{\alpha} g_{\alpha \nu}+\partial_{\nu}\xi_c^{\alpha} g_{\alpha \mu}\,,
\ea
with $\xi_c^{\alpha}=-c\, x^{\alpha}/\l$.
The generalization of the rotation in the 5th dimension ($M_{4\mu})$ is now given by
\be
\delta_v \pi = v_{\mu}x^{\mu}+\partial_{\mu}\pi \( \frac{\l}{2}(e^{2\pi/\l}-1)v^{\mu}+\frac{1}{2\l}v^{\mu}x^2-\frac{1}{\l}(v.x)x^{\mu}\) \, ,
\ee
which after some algebra one may show  corresponds to a gauge transformation of the metric of the form
\be
\xi_v^{\mu}=\( \frac{\l}{2}(e^{2\pi/\l}-1)v^{\mu}+\frac{1}{2\l}v^{\mu}x^2-\frac{1}{\l}(v.x)x^{\mu}\).
\ee
Clearly in the limit $\l \rightarrow \infty$ this reproduces the previous Poincar\'e result.
Similarly the extrinsic curvature
\be
K\mn=-\frac{1}{\sqrt{1+e^{2\pi/\l}(\p \pi)^2}}\(\p_\mu \p_\nu \pi +\frac 1 \l \p_\mu \pi \p_\nu \pi +\frac 1\l g\mn\)
\ee
can be shown to transform as a tensor and one may also check that $e^{-4\pi/\l}$ transforms as a total derivative under both of these transformations and so is the generalization of the tadpole term $\pi$ in the Poincar\'e invariant case.
We are thus led to the most general form of the Lagrangian invariant under the $SO(2,4)$ symmetry to be
\be
\label{LsymAdS}
{\mathcal L} =\sqrt{-g} \, S(g_{\mu\nu},R_{abcd} ,K_{ef}, \nabla_g)+\mathcal{A}(\pi)\,,
\ee
where $\mathcal{A}$ is the generalization of the tadpole term,
\ba
\mathcal{A}=-\frac \l 4 A  (e^{-4\pi/\l}-1)\,.
\ea
Here again, there are two ways in which matter can couple to this theory: Either matter fields couple to the bulk metric $\eta\mn$ and do not transform under the symmetry, or they can be minimally coupled w.r.t. the induced metric,
${\cal L}_m(g_{\mu\nu},\chi^i)$
in which case then should also transform under the symmetry,
\be
\delta_v \chi^i=\xi_v^{\mu}  \partial_{\mu} \chi^i.
\ee
Notice that in this case, the induced metric explicitly depends on $\pi$ (and not uniquely on its derivative), which could provide a natural Vainshtein mechanism.

\subsection{Galileon and Galilean invariance}

The Galileon is defined as the theory invariant under the Galilean contraction of the Poincar\'e group. Intuitively it is defined by taking the limit $(\partial \pi)^2 \ll 1$.
The nonlinear symmetries defined in this limit are
\ba
\delta_c \pi&=& c \\
\delta_v \pi &=& v_{\mu}x^{\mu}.
\ea
Unfortunately there is no non-trivial $\pi$-dependent metric which can be defined for which these symmetries are its gauge transformations. Thus it is not possible to give such a simple expression for the most general invariant Lagrangian. Apart from the tadpole term $\pi$, schematically the relevant terms will be of the form $(\partial \partial \pi)^{n-2}(\partial \pi) (\partial \pi)$, but the form of the contractions is highly nontrivial \cite{Nicolis:2008in}.
However, we can bypass this problem by always defining the possible interactions as the leading non-zero term obtained in the limit $(\partial \pi)^2 \rightarrow 0$ from the Poincar\'e invariant case. This is precisely what we shall do in the following.

\subsection{Conformal Galileon}

The conformal extension of the Galileon, as considered in \cite{Nicolis:2008in}, can be obtained by taking the limit $e^{2\pi/\l}(\partial \pi)^2 \ll 1$ in the Anti-de Sitter case.
This can be done more systematically by redefining $\pi=\l \hat{\pi}$ and then taking the limit $\l \rightarrow 0$.
In this case the form of the infinitesimal nonlinear transformations that survive are
\ba
\delta_c \hat{\pi} &=& c\(1 -x^{\mu}\partial_{\mu} \hat{\pi}\) \\
\delta_v \hat{\pi} &=& v_{\mu}x^{\mu}+\partial_{\mu}\hat{\pi} \( \frac{1}{2}v^{\mu}x^2-(v.x)x^{\mu}\) \, .
\ea
The induced metric becomes $g_{\mu\nu}=e^{-2\hat{\pi}}\eta_{\mu\nu}$ which, modulo a sign convention, is precisely the form used in \cite{Nicolis:2008in} to construct the interactions for this model.
We can similarly determine all the possible interactions by taking the leading non-zero term obtained in the limit $\l \rightarrow 0$ for fixed $\hat{\pi}$. Some care must be taken in this process since it is necessary to construct combinations of invariants which are well-defined in the limit $\l \rightarrow 0$ in order to see all the possible interactions for the conformal Galileon. We discuss how to do this in section~(\ref{conformalgalileon}).

\subsection{Relevant operators and quantum corrections}

\label{qcorrections}

The theories we are considering are clearly in general non-renormalizable, and must be understood as effective field theories.
For each of these 4 classes of theories (Poincar\'e, Anti-de Sitter, Galileon, Conformal Galileon), at first glance one may expect all local operators which are invariant under the appropriate symmetry group to be generated by quantum corrections unless otherwise forbidden by another symmetry. However, if we look at loops from $\pi$ alone, this is not necessarily the case. Indeed, in the Galileon case it has been shown that there is a type of nonrenormalization theorem~\cite{lpr} that stops the leading interactions from receiving loop corrections from $\pi$. This was initially argued in the context of the DGP model, where the simplest nontrivial Galileon model arose, but clearly this argument is independent of the explicit higher dimensional realization. In the present case however we expect loops from matter fields on the brane, \ie matter that couples to the metric $g_{\mu\nu}=\eta_{\mu\nu}+\partial_{\mu}\pi \partial_{\nu}\pi$ to renormalize the intrinsic terms, \ie $\lambda$ and $R$.

Let us first consider the DBI case and write the tension $\lambda= \Lambda_c^4$ and canonically normalize the field $\pi_c =\Lambda_c^2 \pi$ such that
\be
S=\int \d^4x \( -\Lambda_c^4 \sqrt{1+(\partial \pi_c)^2/\Lambda_c^4}+\Lambda_c^4 \).
\ee
Since $\Lambda_c$ is the scale with which na\"ively nonrenormalizable operators are suppressed we might expect the theory to break down when $\partial \pi_c \sim \Lambda_c^2$. However, since the symmetry guarantees that this is the unique operator which does not include higher derivatives, we can still have {\it classical} field configurations for which  $\partial \pi_c \sim \Lambda_c^2$ provided that
all higher derivative combinations which are invariant under the $ISO(1,4)$ symmetry are kept small with respect to $\Lambda_c$ to the power of their engineering dimension. In particular this implies that the extrinsic curvature $K \ll \Lambda_c$, and the intrinsic curvature $R \ll \Lambda_c^2$. In effect, the velocity of the brane may become of order unity provided that the acceleration is much less than unity.
This condition guarantees that the higher order operators, which still respect the DBI symmetry, and come in suppressed by the appropriate power of $\Lambda_c$, nevertheless remain small so that the theory does not go out of control. The condition that quantum fluctuations remain under control is roughly $\sqrt{(\partial \delta \pi_c)^2} \ll \Lambda_c^2$, which is straightforward to satisfy even when $\langle \partial \pi_c \rangle \sim \Lambda_c^2$. In the specific context of inflation this na\"ive argument can be put on a more rigorous footing~\cite{shandera}. In practice the cutoff for the fluctuations depends on the background configuration, and can be determined by looking at the precise form of the suppression of non-renormalizable operators for the perturbations constructed around a given background.

 In the DBI case, the theory was organized in such a way that the unique invariant kinetic term that did not include higher derivatives at the level of the action was taken as the most relevant piece, and all other operators are viewed as irrelevant corrections to this. 
 In the case of IR modified theories the situation is somewhat different. In particular, in theories such as the Galileon, the EFT expansion is further reorganized so that operators which do not give higher derivative equations of motion, but may arise as higher derivatives in the action, are considered to be large (for classical field configurations), \ie as the relevant operators, and all other interactions are viewed as irrelevant corrections. There are only a finite number of such operators, which are protected by the symmetries, and it is this finiteness that is crucial in this being a useful reorganization of the EFT expansion.

The justification of this reorganization is based on several points. Firstly, these are the unique terms that can be taken to be large without immediately introducing ghosts at the scale for which nonlinearities become important, a necessary condition for this to describe a consistent modification of gravity. Secondly, in the Galileon case one can show that there are non-renormalization theorems that prevent the newly defined relevant operators from receiving loop corrections from $\pi$ loops~\cite{lpr}. Thirdly, the real measure of the breakdown of the effective field theory is not determined by derivatives of $\pi_c$, but rather by requiring universally that gradients are small $\partial \ll \Lambda_c$~\cite{Endlich:2010zj,Nicolis:2004qq}. In other words, classically it is possible to have $\partial \partial \pi_c \sim \Lambda_c^3$ and quantum corrections will be small as long as $\partial \ln  (\partial \partial \pi_c)/\Lambda_c \ll 1$. Lastly, and most crucially for phenomenology, the actual cutoff of the EFT may not be the scale that arises in the action $\Lambda_c$, but rather a scale $\tilde{\Lambda}_c$ which takes into account the fact that when looking at perturbations around an arbitrary background configuration $\bar{\pi}_c$, or when coupling to matter on the brane, the kinetic term for the perturbations $\delta \pi_c=\pi_c-\bar{\pi}_c$ gets renormalized by the classical background or matter couplings to be of the form~\cite{Nicolis:2004qq}
\be
S=\int \d^4 x \( -\frac{1}{2}Z^{\mu\nu} \partial_{\mu} \delta \pi_c \partial_{\nu} \delta \pi_c+\dots \).
\ee
The case that has been most studied is the DGP type of coupling, where the Galileon couples to the trace of the matter stress energy through a $\pi T^{\mu}_{\,\, \mu}$ coupling, thus generating a background for the $\pi$ field~\cite{lpr,Nicolis:2004qq}.
If we go deep into the region for which $\partial \partial \bar{\pi}_c \gg \Lambda_c^3$ then the kinetic term renormalization $Z_{\mu\nu} \gg 1$. Thus the perturbations have to be canonically normalized in a manner which depends on the background configuration. For instance for the Galileon theory corresponding to the DGP decoupling limit, after this renormalization the effective scale with which non-renormalizable operators for the perturbations are suppressed is $\tilde{\Lambda}_c\sim \Lambda_c \sqrt{Z}$~\cite{Nicolis:2004qq}, which is much greater than $\Lambda_c$ deep in the so-called strong coupling regime.
This in turn implies that the non-renormalizable interactions are much smaller, and in turn that the quantum corrections are suppressed. This last effect is the main content of the Vainshtein (kinetic chameleon) mechanism \cite{Vainshtein:1972sx}.

Even in the absence of a $\pi T^{\mu}_{\, \,\mu}$ coupling, and without a background for $\pi$, $\bar{\pi}=0$, as soon we have matter on the brane, it couples to the induced metric $g_{\mu\nu}=\eta_{\mu\nu}+\partial_{\mu}\pi\partial_{\nu} \pi$ (this coupling has the virtue of preserving the symmetries unlike the $\pi T^{\mu}_{\,\,\mu}$ coupling), then a kinetic term will be generated in the matter sector and the effective kinetic normalization is
\be
Z^{\mu\nu}=\lambda \eta^{\mu\nu}-T^{\mu\nu} \, ,
\ee
where $\lambda$ is the brane tension. In other words, the effective tension of the brane is $\lambda+ T_{00}$. In a regime where locally $|T_{\mu\nu}| \gg \lambda$ we will also get the Vainshtein mechanism occurring which screens the interactions of the brane modulus.
Notice that in this case the Vainshtein effect is less effective because the kinetic renormalization depends locally on the value of $T_{\mu \nu}$, and so only a small distance from a massive body say, the $\pi$ field no longer feels the suppression effect due to the matter. By contrast, in the more familiar case where there is a background field for $\pi$, the kinetic normalization depends locally on the background field configuration and hence non-locally on the source, and so the screening effect is still felt at some distance from the source, in particular, within the so-called strong coupling radius. Also note that in this case, the absence of ghosts amounts to the condition $\lambda+T_{00} > 0$ which is that the effective tension of the brane is positive semi-definite.\footnote{The same arguments do not apply in the DGP case because of the orbifold projection. This is tied to the fact that $\pi$ has a different physical meaning in the context of the DGP decoupling theory.}

We should stress that whether this resummation, and indeed the precise form of the interactions, is justified from the perspective of its UV completion is a nontrivial question~\cite{Adams:2006sv}, and for our present purposes we ask only that these theories make sense as effective field theories. However, it is worth mentioning that the central question is whether the theory is local/causal/has an analytical S-matrix. The Poincar\'e and Anti-de Sitter theories will be local in higher dimensions, which is arguably all that is really required. From this point of view the galilean limit is somewhat pathological since there is no longer a well-defined metric which is preserved by the symmetry and can be used to give a causal structure. This is consistent with the observation in \cite{Adams:2006sv} that the pure Galileon theory violates the conditions needed for an analytic S-matrix.

Following this logic, it merely behooves us to look for all the possible terms which give rise to second order equations of motion in the actions of the form
\be
S=\int \d^4 x \( \sqrt{-g} \, S(g_{\mu\nu},R_{abcd} ,K_{ef}, \nabla_g)+\mathcal{A}(\pi)\).
\ee
Fortunately this work has already been done for us. All allowed operators must either correspond to Lovelock terms in 4 dimensions or the Gibbons-Hawking-York boundary terms associated with Lovelock terms in 5 dimensions~\cite{Lovelock:1971yv}. The reason is that the Lovelock terms are the unique manifestly covariant terms which do not give rise to higher order equations of motion.
Thus the most general induced action on the probe-brane satisfying our criteria is then
\ba
\label{L}
\L=\sqrt{-g}\(-\lambda+\frac{\mq}{2} R-\mf K-\beta \frac{\mf}{m^2}\mathcal{K}_{\rm GB}\)+\mathcal{A}(\pi)\,,
\ea
where $\mathcal{A}$ is the tadpole contribution, whose exact magnitude depends on the bulk content or additional fields, but whose form is dictated by the symmetries. We have neglected the 4 dimensional Gauss-Bonnet term since it is purely topological, but it is understood that in a higher dimensional realization it is expected to be present.
In this action, indices are raised and lowered with respect to the induced metric $g\mn$, $m$ is the ratio of the 5 dimensional to the 4 dimensional Planck scales, $m=\mf/\mq$,  $K$ is the trace of the extrinsic curvature $K\mn$ on the brane and
the boundary term $\mathcal{K}_{\rm GB}$ is a combination of order extrinsic curvature terms and the induced Einstein tensor \cite{Davis:2002gn}:
\ba
\mathcal{K}_{\rm GB}=-\frac 23 K\mn^3+K K\mn^2-\frac 13 K^3-2G\mn K^{\mu\nu}\,.
\ea
Since the boundary terms $K$ and $\K_{\rm GB}$ arise from the 5 dimensional Lovelock invariant, their projection to the brane is automatically at most second order in derivatives at the level of the equations of motion. The same is not immediately obvious for the induced quantities, such as the Scalar Curvature. To convince ourselves that the equations of motion yield at most 2 derivatives in time, when working in terms of the induced metric $g\mn=q\mn+\p_\mu \pi\p_\nu \pi$, let us first assume $\pi=0$. In that case the restriction to the Lovelock combination ensures that the equations are at most second order in time derivative, and when working in terms of the ADM decomposition, with lapse $N$, $\d s^2=q\mn\d x^\mu\d x^\nu=-N^2\d t^2+\gamma_{ij}(\d x^i +N^i \d t)(\d x^j +N^j \d t)$, the equations are of the schematic form
\ba
\frac{\delta S}{\delta \gamma^{ij}}&=& f_{ij}(\gamma, \dot \gamma, \ddot \gamma, N, \dot N, N^i, \dot N^i)\\
\frac{\delta S}{\delta N}&=& f_{00}(\gamma, \dot \gamma, N^i, \dot N^i)
\ea
where the lapse never enters with more than one time-derivative (here denoted by dotes), and the equation with respect to the lapse propagates a constraint. Space-like derivatives have been omitted from the previous relations for simplicity.

Now, when including the modulus $\pi$, one might worry that the equations of motion could potentially include higher derivative contributions since the induced metric itself contains first order derivatives squared. However, one can always choose the gauge where $\pi=\pi(t)$, and in this gauge $\pi(t)$ enters only through a corresponding redefinition of the lapse to $\tilde N^2=N^2-\dot \pi^2$. Since the equations of motion contained no more than a single derivative of the lapse, it follows trivially that the equations of motion contain no more than second derivatives of $\pi(t)$:
\ba
\frac{\delta S}{\delta \gamma^{ij}}&=& f_{ij}(\gamma, \dot \gamma, \ddot \gamma, N, \dot N, N^i, \dot N^i,\dot \pi, \ddot \pi)\\
\frac{\delta S}{\delta N}&=& f_{00}(\gamma, \dot \gamma, N^i, \dot N^i,\dot \pi, \ddot \pi)
\ea
as do the equations obtained varying with respect to $\pi$:
\ba
\frac{\delta S}{\delta \pi}\sim -\partial_t \(\dot \pi \frac{\delta S}{\delta \tilde N}\)\sim  f_{\pi}(\gamma, \dot \gamma,  \ddot \gamma,  N^i, \dot N^i, \ddot N^i, \dot \pi, \ddot \pi)\,.
\ea
Upon covariantization of these gauge fixed expressions, we clearly see that Lovelock invariants lead to a well-defined Cauchy problem in any gauge.

We will emphasize this fact further by
showing that there is a remarkable set of recursion relations (\ref{dS1}-\ref{dS4}) and (\ref{dS1AdS}-\ref{dS4AdS}) that relate all these different operators, including the 4d Gauss-Bonnet term.
Our aim in the rest of this paper is to show that this simple action is sufficiently general to reproduce all the interactions of the Galileon, Conformal Galileon, their relativistic generalizations, and to determine the correct coupling to gravity.

\section{Probe Brane in Minkowski}
\label{sec:ProbeBrane}

In the higher dimensional bulk the brane is localized at $y=\pi(x^\mu)$, where $y$ parameterizes the direction along the 5th dimension, while the variables $x^\mu$ are 4 dimensional. Assuming the shift $g_{4\mu}$ vanishes in the bulk, the form for the induced metric will then be $g\mn=q\mn+\p_\mu \pi \p_\nu \pi$,
where $q\mn$ is the metric induced on constant-$y$ hypersurfaces.

\subsection{Induced Action}

For the time being, we focus on  the brane position modulus $\pi$ and set $q\mn=\eta\mn$. The generalization to the case with gravity dynamical will be straightforward, and amounts to considering the metric $q\mn$ to be dynamical, as explained in section \ref{sec:Covariant}.
In terms of the brane position modulus, the inverse metric is then
\ba
g^{\mu\nu}=\eta^{\mu\nu}-\frac{\partial^\mu\pi \partial^\nu \pi}{1+(\p \pi)^2}\,,
\ea
where in this section, indices are raised and lowered with respect to the Minkowski metric, unless specified otherwise. The induced extrinsic curvature on the brane is then $K\mn=-\gamma\, \pp $,
where $\gamma$ is the Lorentz factor
\ba
\gamma=\frac{1}{\sqrt{\dbi}} \, .
\ea
The different contributions to the brane action are simply
\ba
S_\lambda \hspace{-5pt}&=&\hspace{-5pt}-\lambda\int \d^4x\sqrt{-g} =-\lambda\int \d^4x\sqrt{\dbi}\\
S_K \hspace{-5pt}&=&\hspace{-5pt}-\mf\int \d^4 x\sqrt{-g}K=\mf \int \d^4x\ \([\Pi]-\gamma^2 [\phi]\)\\
S_R \hspace{-5pt} &=& \hspace{-5pt} \frac{\mq}{2}\int \d^4 x\sqrt{-g}R=
\frac{\mq}{2} \int \d^4x\, \gamma\(  \([\Pi]^2-[\Pi^2]\)+2 \gamma^{2}\([\phi^2]-[\Pi] [\phi]\)\)\hspace{20pt}
\ea
and
\ba
S_{GB}\hspace{-5pt}&=&\hspace{-5pt}-\beta \frac{\mf}{m^2}\int \d^4x \sqrt{-g}\, \K_{\rm GB}\nn\\
\hspace{-5pt}&=&\hspace{-5pt}\beta \frac{\mf}{m^2}\int \d^4x\, \gamma^2 \Big(
\frac 23 \([\Pi]^3+2[\Pi^3]-3[\Pi][\Pi^2]\)+4\gamma^2 ([\Pi][\phi^2]-[\phi^3])\\
\hspace{-5pt} && \hspace{-5pt}\phantom{\beta \frac{\mf}{m^2}\int \d^4x \Big(}
-2\gamma^2([\Pi]^2-[\Pi^2])[\phi]
\Big)\,,\nn
\ea
where we used the same notation as \cite{Nicolis:2008in}, with $\Pi\mn=\pp$ and square brackets $[...]$ represent the trace (w.r.t. $\eta_{\mu\nu}$) of a tensor. Furthermore we also introduced the following notation
\ba
[\phi^n]\equiv\p \pi \,.\, \Pi^n \,.\, \p \pi\,,
\ea
so in particular $[\phi]=\p_\mu \p_\nu\pi \p^\mu \pi \p^\nu \pi$.

The higher dimensional origin of these terms ensures that equations of motion remain at most second order in derivatives, and before providing the exact equation of motion for the field $\pi$, it is worth pointing out the following recursive relations that exist between each term in the induced action:
\ba
\label{dS1}
&&\frac{\delta}{\delta \pi}\( \sqrt{-g}\)=K\\
\label{dS2}
&&\frac{\delta}{\delta \pi} \(\sqrt{-g}K\)=R\\
\label{dS3}
&&\frac{\delta}{\delta \pi} \(\sqrt{-g}R\)=\frac32 \K_{\rm GB}\\
\label{dS4}
&&\frac{\delta}{\delta \pi} \( \sqrt{-g}\K_{\rm GB}\)=\frac23 \L_{GB_4}=
\frac 23\(R^2-4R\mn^2+R_{\mu\nu\alpha\beta}^2\)\,,
\ea
where in the last expression, we see appearing the four-dimensional Gauss-Bonnet term at the level of the equations of motion.
Here $\frac{\delta}{\delta \pi}$ is the Euler-Lagrange derivative
\be
\frac{\delta}{\delta \pi}=\frac{\partial}{\partial \pi}-\partial_{\mu} \frac{\partial}{\partial \partial_{\mu}\pi}+\partial_{\mu} \partial_{\nu}\frac{\partial}{\partial \partial_{\mu} \partial_{\nu}\pi}\,.
\ee
 Notice that in four dimensions, the Gauss-Bonnet term is topological at the level of the action, but at the level of the equations of motion, as it appears here, it produces a non-trivial, yet ghost-free contribution.
As we will see in section \ref{sec:AdS}, this important recursive relation remains valid around any maximally symmetric background, and will actually only be very slightly modified around an arbitrary background.

First of all, this recursive relation, makes it very clear that the equations of motion remain second order in derivatives and the theory is therefore free of any manifest ghost-like instability (of course there may exist configurations around which perturbations are ghostly despite having a well-defined Cauchy problem. However, such configurations invariably lie outside of the regime of validity of the effective field theory). Furthermore, this provides an explicit framework to derive the entire Lovelock family of higher-order terms
in arbitrary dimensions, \cite{Lovelock:1971yv}.

Using these relations, the modified Klein-Gordon equation is then expressed as follows
\ba
\frac{\delta {\cal L}}{\delta \pi}\hspace{-5pt}&=&\hspace{-5pt}\lambda \, \gamma \, \( [\Pi]-\gamma^2[\phi]\)
-
\mf \, \( \gamma^2\([\Pi]^2-[\Pi^2]\)+2 \gamma^2\, \([\phi^2]-[\Pi][\phi]\) \)\\
\hspace{-5pt}&-&\hspace{-5pt}\frac{\mq}{2}\gamma^3\Big([\Pi]^3+2[\Pi^2]-3[\Pi][\Pi^2]
+3\gamma^2\(2([\Pi][\phi^2]-[\phi^3])-([\Pi]^2-[\Pi^2])[\phi]\)\Big)\nn\\
\hspace{-5pt}&-&\hspace{-5pt}\frac{2}{3}\beta \frac{\mf}{m^2}\gamma^6\Big(
[\Pi]^4-6[\Pi]^2[\Pi^2]+8[\Pi][\Pi^3]+3[\Pi^2]^2-6[\Pi^4]
\Big)
=0\,,\nn
\ea
which is a complete generalization of the DBI model (first term), and the Galileon contributions as we shall below. It is worth pointing out that, for classical field configurations, all the terms in these equations of motion can be similarly significant without running out of control from higher order contributions, as a consequence of the arguments of section \ref{qcorrections}.

\subsection{Recovering the Galileon}
As promised, in the non-relativistic limit $(\p \pi)^2\ll 1$, we recover for free the different Galileon contributions as derived by hand in \cite{Nicolis:2008in}:
\ba
S_2\hspace{-5pt}&=&\hspace{-5pt}S_\lambda^{NR}=-\frac{\lambda}{2}\int\d^4 x\, (\p \pi)^2\\
S_3\hspace{-5pt}&=&\hspace{-5pt}S_K^{NR}=\frac{\mf}{2}\int\d^4 x\,  (\p \pi)^2\, \Box \pi\\
S_4\hspace{-5pt}&=&\hspace{-5pt}S_R^{NR}=\frac{\mq}{4}\, \int \d^4 x\, (\p \pi)^2\, \((\Box \pi)^2-(\partial_\mu \partial_\nu \pi)^2\)\label{S4nr}\\
S_5\hspace{-5pt}&=&\hspace{-5pt}S_{GB}^{NR}=\beta\frac{\mf}{3m^2}\int \d^4 x\,
(\p \pi)^2 \((\Box \pi)^3+2 (\partial_\mu \partial_\nu\pi)^3-3 \Box \pi  (\partial_\mu \partial_\nu\pi)^2\)\,.
\label{S5nr}
\ea
Notice that $S_4$ and $S_5$ take a slightly different form than \cite{Nicolis:2008in}, but the expressions are actually equivalent after noticing that the following combinations are total derivative in Minkowski, as mentioned in \cite{Deffayet:2009wt}:
\ba
\L^{div}_4\hspace{-5pt}&=&\hspace{-5pt}2([\Pi][\phi]-[\phi^2])+([\Pi]^2-[\Pi^2])(\p \pi)^2\\
\L^{div}_5\hspace{-5pt}&=&\hspace{-5pt}2([\Pi]^2-[\Pi^2])[\phi]-4([\Pi][\phi^2]-[\phi^3])+([\Pi]^3+2[\Pi^3]-3[\Pi][\Pi^2])(\p \pi^2)\,.
\ea
Our derivation has now shed light on one crucial aspect of the Galileon theory that always appeared remarkable: How could higher derivative operators in the Lagrangian give rise to conventional second order equations of motion? We now see that it is precisely because they follow from the special Lovelock combinations. It is the topological nature of these terms in lower dimensions which is connected to their equations of motion being second order. This can be seen through the fact that the expression for these invariants requires the use of the antisymmetric Levi-Civita symbol, and as shown in \cite{Deffayet:2009mn} this is the crucial property to demonstrating the well-defined Cauchy problem.

\section{Probe brane in Anti-de Sitter}
\label{sec:AdS}


\subsection{Induced Action}

In general, the bulk contains a cosmological constant which induces a five-dimensional de Sitter or Anti-de Sitter (AdS) warped geometry. For definiteness, we choose an AdS bulk, of the form
\ba
\d s^2=\d y^2 +e^{-2 y/\l} \eta\mn \d x^\mu \d x^\nu\,,
\ea
where the AdS length $\l$ is related to the bulk cosmological constant $\Lambda$ via the 5d Planck scale $M_5$ and the coupling $\beta$ to the Gauss-Bonnet terms:
\ba
\Lambda=-\frac{6 \mf}{\l^2}\(1-\frac{2\beta}{\l^2 m^2}\)\,.
\ea
In this section both the covariant derivative and the raising and lowering of indices are still performed using the Minkowski metric.
In terms of the Lorentz factor $\tg$,
\ba
\tg=\frac{1}{\sqrt{1+e^{2\pi/\l}(\p \pi)^2}}\,,
\ea
the induced metric and extrinsic curvature on the brane of position $y=\pi(x^\mu)$ are
\ba
g\mn&=&e^{-2\pi/\l}\eta\mn+\p_\mu \pi \p_\nu \pi\\
K\mn&=&-\tg\(\p_\mu \p_\nu \pi +\frac 1 \l \p_\mu \pi \p_\nu \pi +\frac 1\l g\mn\)\\
K&=& -e^{2\pi/\l}\tg
\(\Box \pi-\tg ^2 e^{2\pi/\l}[\phi]+\frac {\tg^2}{\l} (\p \pi)^2+\frac 4 \l e^{-2\pi/\l}\)\,.
\label{KAdS}
\ea
The expression for the induced scalar curvature is
\ba
\label{RAdS}
R&=&\tilde \gamma^4 e^{4\pi/\l}\Bigg[
\tilde\gamma^{-2}([\Pi]^2-[\Pi^2])+2 e^{2\pi/\l}([\phi^2]-[\Pi][\phi])\\
&&-\frac{6}{\l^2}e^{-2\pi/\l}(\p \pi)^2\(1+2 e^{2\pi/\l}(\p \pi)^2\)-\frac 8 \l [\phi]
+\frac 2 \l e^{-2\pi/\l}\(3+4e^{2\pi/\l}(\p \pi)^2\)[\Pi]
\Bigg] \, ,\nn
\ea
and finally that of the Gauss-Bonnet boundary term is
\ba
\label{KGBAdS}
\K_{\rm GB}\hspace{-5pt}&=&\hspace{-5pt}-\tg^5 e^{6\pi/\l}\Bigg[
\frac 2{3\tg^2} \([\Pi]^3-3[\Pi][\Pi^2]+2[\Pi^3]\)]
\\ \hspace{-5pt}&&\hspace{-5pt}
+2e^{2\pi/\l}\(2([\Pi][\phi^2]-[\phi^3])-([\Pi]^2-[\Pi^2])[\phi]\)\nn\\
\hspace{-5pt}&&\hspace{-5pt} -\frac{12}{\l}([\Pi][\phi]-[\phi^2])+\frac 2 \l e^{-2\pi/\l}(2+3 e^{2\pi/\l}(\p \pi)^2)([\Pi]^2-[\Pi^2])\nn\\
\hspace{-5pt}&&\hspace{-5pt}+\frac{2}{\l^2}e^{-2\pi/\l}(-7+3e^{2\pi/\l}(\p \pi)^2)[\phi]
\nn\\ \hspace{-5pt}&&\hspace{-5pt}
+\frac{2}{\l^2}e^{-4\pi/\l}(3+4e^{2\pi/\l}(\p \pi)^2-3 e^{4\pi/\l}(\p \pi)^4)[\Pi]\nn \hspace{-100pt}\\
\hspace{-5pt}&&\hspace{-5pt}-\frac{2}{\l^3}e^{-6\pi/\l}(4+13e^{2\pi/\l}(\p \pi)^2+15 e^{4\pi/\l}(\p \pi)^4)
\Bigg]\,. \nn
\ea
The induced action on the brane \eqref{L} then provides a generalization of the DBI/Galileon action to a family of models, where the analogue of the recursive relations (\ref{dS1}-\ref{dS4}) remain valid and give rise to the following very simple relations despite the complicated expressions in \eqref{RAdS} and \eqref{KGBAdS}:
\ba
\label{dS1AdS}
&&\frac{\delta}{\delta \pi}\( \sqrt{-g}\)=e^{-4\pi/\l}K\\
\label{dS2AdS}
&&\frac{\delta}{\delta \pi} \(\sqrt{-g}K\)=e^{-4\pi/\l}\(R+\frac{16}{\l^2}\)\\
\label{dS3AdS}
&&\frac{\delta}{\delta \pi} \(\sqrt{-g}R\)=e^{-4\pi/\l}\(\frac32 \K_{\rm GB}+\frac{3}{\l^2}K\)\\
\label{dS4AdS}
&&\frac{\delta}{\delta \pi} \(\sqrt{-g}\, \K_{\rm GB}\)=e^{-4\pi/\l}\(\frac23\L_{GB_4}+\frac 2{3\l^2} R-\frac{32}{\l^4}\)\,.
\ea
Here $R$ and $\L_{GB_4}$ are respectively the Ricci scalar and Gauss-Bonnet scalar with respect to the induced metric $g\mn$.

\subsection{Recovering DBI}
The DBI limit of this theory is trivial to obtain, and simply amounts to taking the { conventional} limit of the probe-brane action \eqref{L} in the AdS case, where the action is dominated by the lowest order operators, the tension and the tadpole contribution, which takes the form
\ba
\mathcal{A}(\pi)=\lambda e^{-4\pi/\l}\,,
\ea
while all subsequent contributions are treated as small, or not present. More concretely this amounts to taking the corner of phase space where $M_4,M_5\to 0$, so that the resulting action in this limit is
\ba
\label{DBIL}
\L_{\rm DBI}&=&-\sqrt{-g}\lambda+\lambda e^{-4\pi/\l}\nn\\
&=&e^{-4\pi/\l}\(-\lambda \sqrt{1+e^{2\pi/\l}(\p \pi)^2}+\lambda\)\,.
\ea
Notice that in this limit, the main contribution to the tadpole term $\mathcal{A}$ arises from Chern-Simon terms in the DBI action. This is different to the static Randall-Sundrum brane, for which $M_4\to 0$ while $M_5$ remains finite. In this case the tadpole comes from integrating the 5 dimensional curvature across the bulk giving  $\mathcal{A}=2 \mf/\l \, e^{-4\pi/\l}$, while the Isra\"el matching condition imposes $\lambda=6\mf /\l$. Both of these contributions are then canceled to leading order in $(\p \pi)^2$ by the extrinsic curvature term, $K=-4 \tilde \gamma /\l$.

\subsection{Recovering the Conformal Galileon}
\label{conformalgalileon}

The conformal Galileon, on the other hand is obtained by taking the non-relativistic limit of AdS, where $e^{2\pi/\l}(\p \pi)^2\ll 1$. This can be achieved more systematically by defining $\pi=\l \hat{\pi}$ and then taking the limit $\l \rightarrow 0$. However as we now explain this must be done with some care. It is first necessary to construct specific combinations of invariants to pick out the desired interactions. For the tadpole term we clearly obtain in the limit
\be
\L_1= \lim_{\l \rightarrow 0} e^{-4{\pi}/\l}=e^{-4\hat{\pi}}
\ee
which remains invariant. For the tension we must look at the combination
\be
\L_2 =\lim_{\l \rightarrow 0} \frac{1}{\l^2}\( -\sqrt{-g}+e^{-4\hat{\pi}}\)=-\frac{1}{2} e^{-2 \hat{\pi}}(\partial \hat{\pi})^2\,.
\ee
For the extrinsic curvature we must look at the combination
\ba
\L_3 &=& \lim_{\l \rightarrow 0} \frac{1}{\l^3}\(-\sqrt{-g} K +\frac{2}{\l}e^{-4\pi/\l}-\frac{6}{\l}\sqrt{-g}  \)\\
&=& \frac{1}{2}(\partial \hat{\pi})^2 \Box \hat{\pi}-\frac{1}{4} (\partial \hat{\pi})^4 + \text{total derivative} \, .
\ea
Modulo the convention difference ($\hat{\pi}_{\rm here} \rightarrow -\pi_{\rm there}$) this is precisely the conformally invariant term derived by different means in \cite{Nicolis:2008in}.
Proceeding in this way we may construct all the interactions of the conformal Galileon. In particular, the two remaining interactions that give second order equations of motion are given by
\ba
\L_4&=&\lim_{\l\to 0}\frac{\sqrt{-g}}{\l^4} \(R-\frac{6}{\l}K+\frac{24}{\l^2}\)\\
&=&\frac{1}{20}e^{2\hat \pi}(\p \hat \pi)^2
\(10([\hat \Pi]^2-[\hat \Pi^2])+4((\p \hat \pi)^2 \Box \hat \pi-[\hat \phi])+3(\p \hat \pi)^4\)\nn\\
&&+\text{ total derivative}\,,\nn
\ea
and
\ba
\L_5&=&\lim_{\l\to 0}\frac{\sqrt{-g}}{\l^5} \(-\K_{\rm GB}-\frac{8}{3\l}R-\frac{10}{\l^2}K-\frac{32}{\l^3}\)\\
&=&e^{4\hat \pi} (\p \hat \pi)^2
\Big[\frac13([\hat \Pi]^3+2[\hat \Pi^3]-3[\hat \Pi][\hat \Pi^2])+(\p \hat \pi)^2([\hat \Pi]^2-[\hat \Pi^2])\nn\\
&&+\frac{10}{7}(\p \hat \pi)^2((\p \hat \pi)^2[\hat \Pi]-[\hat \phi])-\frac{1}{28}(\p \hat \pi)^6\Big]
+\text{total derivative}\,.\nn
\ea
Unlike the pure Galileon case, this limit is remarkable in that the symmetry group remains intact as $SO(2,4)$ despite the fact that we are taking a type of non-relativistic limit. This is somewhat analogous to the fact that the low energy theory of a D3-brane corresponds to a (super)Yang-Mills theory which still classically preserves the same conformal symmetry which was inherited from higher dimensional isometry group.

\section{Coupling to gravity}
\label{sec:Covariant}

\subsection{Induced Action}

For simplicity we have only considered so far a Minkowski or AdS background geometry for the modulus field $\pi$. The coupling to gravity is however straightforward and only amounts to generalizing this approach to an arbitrary 4 dimensional metric $q\mn$, so that the induced is given by
\ba
g\mn=q\mn(x)+\p_\mu\pi \p_\nu \pi\,.
\ea
The generalization of the contributions to the induced metric are then simply
\ba
K\hspace{-5pt}&=&\hspace{-5pt}-\gamma\([\Pi]-\gamma^2 [\phi]\)\\
R\hspace{-5pt}&=&\hspace{-5pt}\bar R-2\gamma^2\p^\mu \pi \p^\nu \pi\, \bar R\mn+\gamma^2\([\Pi]^2-[\Pi^2]\)-2\gamma^4\([\Pi][\phi]-[\phi^2]\)
\ea
and
\ba
\label{L5quv}
\K_{\rm GB}\hspace{-5pt}&=&\hspace{-5pt}-\gamma^3\Bigg[
\frac 23 \([\Pi]^3+2[\Pi^3]-3[\Pi][\Pi^2]\)
+4\gamma^2([\Pi][\phi^2]-[\phi^3])
\\ \hspace{-5pt}&&\hspace{5pt}
-2\gamma^2([\Pi]^2-[\Pi^2])[\phi]
+\p^\mu\pi \p ^\nu \pi\(4\bar R_{\mu \alpha}\Pi^{\alpha}_{\, \nu}-2\bar R\mn [\Pi]\)
\nn\\ \hspace{-5pt}&&\hspace{5pt}
+2\bar R_{\alpha \mu \beta\nu}\Pi^{\mu\nu} \p^\alpha \pi \p ^\beta \pi
-[\phi]\bar R
-2 \bar G\mn \Pi^{\mu\nu}
\Bigg]\,,\nn
\ea
where $R$ is the Ricci scalar induced on the brane, $R=R[g\mn]$ with induced metric $g\mn=q\mn+\p_\mu\pi\p_\nu \pi$ while bar quantities are computed with respect to the metric $q\mn$, $\bar R=R[q\mn]$, and similarly for the Riemann, Ricci and Einstein tensors.
Moreover, in the previous expression, indices are raised and lowered using the metric $q\mn$, and the covariant derivatives are taken with respect to $q\mn$. In particular, $\Pi\mn=\nabla_\mu \nabla_\nu \pi$ and similarly for $\phi$.

\subsection{Recovering the Covariant Galileon}

As pointed out in  \cite{Deffayet:2009wt, Deffayet:2009mn}, when studying the Galileon around a generic background metric $q\mn$, the actions \eqref{S4nr} and \eqref{S5nr} involve more
than two derivatives at the level of the equations of motion a non-minimal coupling needs to be introduced to ensure the stability of the theory. As we shall see below, such corrections are automatic in this formalism, and when working around a generic background,
and in the non-relativistic limit, the expression for $S_4$ is then
\ba
S_4\hspace{-5pt}&=&\hspace{-5pt}\frac{\mq}{2}\int \d^4 x \sqrt{-g} R\nn \\
\hspace{-5pt}&=&\hspace{-5pt}\frac{\mq}{2} \int \d^4 x\sqrt{-q}\Bigg[
\bar R \(1+\frac 12 (\p \pi)^2-\frac 18 (\p \pi)^4\)-2 \p^\mu \pi \p^\nu \pi \bar R\mn \(1-\frac 12 (\p \pi)^2\)\nn\\
\hspace{-5pt}&&\hspace{-5pt}\phantom{\mq \int \d^4 x\sqrt{-q}\Bigg[}
-\frac 12 \([\Pi]^2-[\Pi^2]\)(\p \pi)^2-2 \([\Pi][\phi]-[\phi^2]\)
\Bigg]\,,
\ea
so that after canonically normalizing the modulus $\hat \pi=m M_4 \, \pi$,
the terms scaling as $\Lambda_3^{6}$
are precisely that derived by Deffayet {\it et.al.} in \cite{Deffayet:2009wt, Deffayet:2009mn}:
\ba
S_{4}^{dec}=\frac{1}{4\Lambda_3^6}\int \d^4x\sqrt{-q}\(([\hat \Pi]^2-[\hat \Pi^2])(\p \hat \pi)^2-\frac 14 \bar R (\p \hat \pi)^4\)\,.
\ea
Notice however that this framework has the power to provide the complete covariant generalization beyond the non-relativistic and decoupling limit.

Similarly, the covariant generalization of the final term \eqref{S5nr} can be derived working with the generic metric $q\mn$ and is provided in \eqref{L5quv},
so that after canonical normalization, and upon integration by parts, the terms scaling as $\Lambda_3^{9}$ in the non-relativistic limit are once again precisely as determined in \cite{Deffayet:2009wt, Deffayet:2009mn}:
\ba
S_{5}^{dec}\hspace{-5pt}&=&\hspace{-5pt}\frac{\beta}{\Lambda_3^9}\int \d^4x\sqrt{-q}\Big[
-\frac 23 \([\hat \Pi]^3+2[\hat \Pi^3]-3[\hat \Pi][\hat \Pi^2]\)(\p \hat \pi)^2
+4([\hat \Pi][\hat \phi^2]-[\hat \phi^3])\\
\hspace{-5pt}&&\hspace{-5pt}\phantom{\frac{1}{2\Lambda_3^9}\int \d^4x\sqrt{-q}\Big[}
-2([\hat \Pi]^2-[\hat \Pi^2])[\hat \phi]
-\p^\mu\hat \pi \p ^\nu \hat \pi\(4\bar R_{\mu \alpha}\hat \Pi^{\alpha}_{\, \nu}-2\bar R\mn [\hat \Pi]\)(\p \hat \pi)^2
\nn\\ \hspace{-5pt}&&\hspace{-5pt}\phantom{\frac{1}{2\Lambda_3^9}\int \d^4x\sqrt{-q}\Big[}
-2\bar R_{\alpha \mu \beta\nu}\hat \Pi^{\mu\nu} \p^\alpha \hat \pi \p ^\beta \hat \pi (\p \hat \pi)^2+[\hat \phi]\bar R (\p \hat \pi)^2
\Big]\nn \\
\hspace{-5pt}&=&\hspace{-5pt}\frac{\beta}{\Lambda_3^9}\int \d^4x\sqrt{-q} (\p \hat \pi)^2
\Big[\frac 13 \([\hat \Pi]^3+2[\hat \Pi^3]-3[\hat \Pi][\hat \Pi^2]\) -2 \bar G\mn \hat \Pi^{\mu \alpha}\p_\alpha \p^\nu \hat \pi\Big]\,.
\ea

\subsection{Higher dimensional picture}

The simplicity with which we have derived a consistent theory which couples to gravity and is purely 4 dimensional makes one wonder whether it is necessary to go the higher dimensional picture for anything other than guidance in how to construct the 4 dimensional theory. The reason it is ultimately necessary is that gravity necessary breaks the nonlinearly realized symmetries, and so we expect at least $1/M_4^2$ suppressed corrections which do not respect the symmetries. Classically this is not a problem, but at the quantum level we expect to generate potentially dangerous operators which break the low energy symmetry. For instance if the cutoff is $\Lambda$ we expect to generate a mass for $\pi$ of order $m_{\pi}^2 \sim \Lambda^4/M_{\rm pl}^2$. This may not be a problem as long as the cutoff is taken to be well below the Planck scale, however in general if the symmetry is not present in the UV completion, there appears to be nothing to protect the mass from becoming Planckian in the UV, which would certainly render the low energy effective theory inconsistent. Indeed there are now standard arguments which show that all continuous global symmetries must be broken in a theory of quantum gravity, and that this breaking is at the Planck scale.

This fact was obvious since the nonlinear symmetries did not commute with the generators of the 4 dimensional Poincar\'e group. As soon as we make the 4 dimensional coordinate transformations local, it is no longer possible for the extra symmetries to act in the same way. The solution to this problem is to promote the whole symmetry group to a local one. In other words replace the Poincar\'e or Anti-de Sitter symmetries with nonlinearly realized 5 dimensional diffeomorphisms. Since the symmetry is now local, it evades the arguments about being broken at the Planck scale. This is of course precisely what we get from the 5 dimensional picture, where the brane is the object which is manifestly invariant under 4d diffeomorphisms and nonlinearly realizes the 5d diffeomorphisms, and is also what happens in the DGP construction. From this perspective then, the Covariant Galileon model could only make sense as a limit of some higher dimensional braneworld construction.

Implicit in this construction is that matter should couple in a manner which is invariant under 5d diffeomorphisms. This is automatic for matter that either lives in the bulk and couples to the bulk metric, or lives on the brane and couples to the brane metric. To exhibit the Vainshtein mechanism in its usual form, it is necessary that in some limit matter couples linearly to $\pi$. In DGP, this coupling arises naturally because the brane metric kinetic term mixes with the $\pi$ kinetic term, and so upon diagonalization one picks up the appropriate conformal coupling. In the probe brane limit with a Minkowski bulk, couplings of this type do not arise. However, they are present for the anti-de Sitter bulk because the induced metric depends explicitly on $\pi$ through the warping, and indeed in the Conformal Galileon limit, the induced metric is just conformally related to the bulk metric, and so matter will naturally pick up the appropriate conformal coupling.

Fortunately it is now transparent how to realize the most general form of these theories by taking an action which is a sum of 4d Lovelock invariants localized on the brane, and 5d Lovelock invariants in the bulk (and their associated boundary terms) with potentially different coefficients on each side of the brane, including in general a bulk cosmological constant (which may be different on each side of the brane).
There exists a decoupling limit where we recover the Minkowski or AdS DBI/Galileon effective field theories. 
For example, a limit corresponds to taking the coefficients of the bulk Lovelock invariants to infinity (decoupling bulk gravity) in such a way that the bulk AdS curvature remains finite and the difference of the Lovelocks on each side of the brane remains finite, so that their boundary values still contribute to the brane action.

Note that crucially this probe brane decoupling limit will in general not be the same as the DGP decoupling limit which is appropriate for tensionless orbifold branes since in the above we rely on the presence of a brane tension to generate a kinetic term for the brane modulus $\pi$. Notice also that the volume of the extra dimensions may be finite. This does not preclude these being possible IR modifications of gravity, if the volume is still much larger than the Hubble scale. It merely indicates that there will be a transition from a 4d regime dominated by the brane induced curvature, to a 5d regime and back to a 4d regime dominated by the Kaluza-Klein zero mode as we flow into the IR. It is plausible that, when an analysis similar to the DGP decoupling limit derivation~\cite{lpr} is performed for theories of this type where the brane is an orbifold plane, we can recover similar results even in the zero tension case. However, there does appear to be a problem since as pointed out in Ref.~\cite{Nicolis:2008in} in the absence of a tension, even with all the Lovelock invariants there do not appear to be sufficient terms to reproduce all the Galileon interactions. The connection between the different limits, and fully understanding the higher dimensional framework is left to future study.

\section{Conclusions}
\label{sec:Conclusion}

The aim of this manuscript is to emphasize the profound relation existing between different types of higher derivative 4 dimensional theories that possess a well defined Cauchy problem and exhibit special symmetries. The connection between these theories relies on the nonlinear realization of an underlying higher dimensional symmetry.
In doing so, we have formulated
a general class of effective field theories which, in different limits, reproduce the Galileon, covariant Galileon and its conformal extension as well as the familiar DBI type of Lagrangians. These theories are a new class of IR modified gravities which have a clear interpretation in extra dimensions, and may exhibit the same interesting phenomenology, such as the Vainshtein mechanism, that has now become familiar from the DGP model.

Whilst we have focused here on the formulation of this theory and the connection between different limits, the physical phenomenology of this new framework is vast and is left for further investigations. The extension of DBI for instance should be studied with care and could lead to a rich phenomenology in the context of DBI inflation. In the DGP limit, the presence of the extra mode modifies structure formation, precision solar system tests, binary pulsars dynamics and many other physical observables. We have provided here the tools to extend these tests not only to the Galileon and Conformal Galileon model, but to a much broader class of theory that connects with DBI.

Finally, we have argued that since such theories are non-renormalizable, they should be understood as effective field theories, valid below some cutoff scale $\Lambda_c$. Fully understanding whether these theories can make sense at the quantum level is beyond the scope of this article, however since these models are contiguous with the Galileon effective theories, we expect to rely on the same arguments use there, as discussed in section~(\ref{qcorrections}) to make sense of them. These arguments rely on demonstrating that there can exist classical field configurations, where fields become na\"ively of the order of the cutoff $\Lambda_c$, but quantum fluctuations around them still remain small, provided that derivatives are small $\partial \ll \Lambda_c$, see \cite{lpr, Nicolis:2004qq, Endlich:2010zj}. Furthermore, in the presence of a source the actual cutoff can be raised to a much larger scale $\tilde{\Lambda}_c$ via the Vainshtein/kinetic chameleon mechanism, a necessary requirement for theories of this type to successfully confront phenomenology. We leave the exploration of these details to future work.

\section*{\small Acknowledgments}

We would like to thank Clare Burrage, Cedric Deffayet, Stanley Deser, Lavinia Heisenberg, Kurt Hinterbichler, Justin Khoury, David Seery and Sarah Shandera for useful comments. The work of CdR is supported by the SNF and that of AJT at the Perimeter Institute is supported in part by the Government of Canada through NSERC and by the Province of Ontario through MRI.

\pagebreak


\begin{thebibliography}{99}





\bibitem{DGP}
  G.~R.~Dvali, G.~Gabadadze and M.~Porrati,
  Phys.\ Lett.\ B {\bf 485}, 208 (2000)
  [arXiv:hep-th/0005016];
  G.~R.~Dvali and G.~Gabadadze,
  Phys.\ Rev.\  D {\bf 63}, 065007 (2001)
  [arXiv:hep-th/0008054].
%
%
%
\bibitem{gigashif}
  G.~Gabadadze and M.~Shifman,
  Phys.\ Rev.\  D {\bf 69}, 124032 (2004)
  [arXiv:hep-th/0312289].
%
 \bibitem{cascading}
 C.~de Rham, G.~Dvali, S.~Hofmann, J.~Khoury, O.~Pujolas, M.~Redi and A.~J.~Tolley,
  Phys.\ Rev.\ Lett.\  {\bf 100}, 251603 (2008)
  [arXiv:0711.2072 [hep-th]];
  C.~de Rham, S.~Hofmann, J.~Khoury and A.~J.~Tolley,
  JCAP {\bf 0802}, 011 (2008)
  [arXiv:0712.2821 [hep-th]];
C.~de Rham, J.~Khoury and A.~J.~Tolley,
  Phys.\ Rev.\ Lett.\  {\bf 103}, 161601 (2009)
  [arXiv:0907.0473 [hep-th]];
  C.~de Rham, J.~Khoury and A.~J.~Tolley,
  arXiv:1002.1075 [hep-th].
%
\bibitem{aux1}
  G.~Gabadadze,
  Phys.\ Lett.\  B {\bf 681}, 89 (2009)
  [arXiv:0908.1112 [hep-th]];
  C.~de Rham,
  Phys.\ Lett.\  B {\bf 688}, 137 (2010)
  [arXiv:0910.5474 [hep-th]].

%
%
  \bibitem{intersecting}
 O.~Corradini, K.~Koyama and G.~Tasinato,
  Phys.\ Rev.\  D {\bf 78}, 124002 (2008)
  [arXiv:0803.1850 [hep-th]];
  Phys.\ Rev.\  D {\bf 77}, 084006 (2008)
  [arXiv:0712.0385 [hep-th]].
%
\bibitem{Kaloper:2007ap}
  N.~Kaloper and D.~Kiley,
  JHEP {\bf 0705}, 045 (2007)
  [arXiv:hep-th/0703190];
  N.~Kaloper,
  Mod.\ Phys.\ Lett.\  A {\bf 23}, 781 (2008)
  [arXiv:0711.3210 [hep-th]].
%
%
%
%
%
\bibitem{Silverstein:2003hf}
  E.~Silverstein and D.~Tong,
  Phys.\ Rev.\  D {\bf 70}, 103505 (2004)
  [arXiv:hep-th/0310221];
%
  M.~Alishahiha, E.~Silverstein and D.~Tong,
  Phys.\ Rev.\  D {\bf 70}, 123505 (2004)
  [arXiv:hep-th/0404084];
%
  D.~Tong,
{\it  In *Tsukuba 2004, SUSY 2004* 841-844}.
%
%
%
%
%
%
\bibitem{nima}
  N.~Arkani-Hamed, H.~Georgi and M.~D.~Schwartz,
  Annals Phys.\  {\bf 305}, 96 (2003)
  [arXiv:hep-th/0210184];
%
%
\bibitem{lpr}
 M.~A.~Luty, M.~Porrati and R.~Rattazzi,
  JHEP {\bf 0309} (2003) 029
  [arXiv:hep-th/0303116];
%
  M.~Porrati and J.~W.~Rombouts,
  Phys.\ Rev.\  D {\bf 69}, 122003 (2004)
  [arXiv:hep-th/0401211];
%
%
%
\bibitem{Aharony:1999ti}
  O.~Aharony, S.~S.~Gubser, J.~M.~Maldacena, H.~Ooguri and Y.~Oz,
  Phys.\ Rept.\  {\bf 323} (2000) 183
  [arXiv:hep-th/9905111].
%
%
%
\bibitem{Nicolis:2008in}
  A.~Nicolis, R.~Rattazzi and E.~Trincherini,
  Phys.\ Rev.\  D {\bf 79}, 064036 (2009)
  [arXiv:0811.2197 [hep-th]].
%
\bibitem{Deffayet:2009wt}
  C.~Deffayet, G.~Esposito-Farese and A.~Vikman,
  Phys.\ Rev.\  D {\bf 79}, 084003 (2009)
  [arXiv:0901.1314 [hep-th]];

\bibitem{Deffayet:2009mn}
  C.~Deffayet, S.~Deser and G.~Esposito-Farese,
  Phys.\ Rev.\  D {\bf 80}, 064015 (2009)
  [arXiv:0906.1967 [gr-qc]].
%
%
%
\bibitem{Chow:2009fm}
  N.~Chow and J.~Khoury,
  Phys.\ Rev.\  D {\bf 80}, 024037 (2009)
  [arXiv:0905.1325 [hep-th]];
%
  F.~P.~Silva and K.~Koyama,
  Phys.\ Rev.\  D {\bf 80}, 121301 (2009)
  [arXiv:0909.4538 [astro-ph.CO]];
%
  T.~Kobayashi,
  arXiv:1003.3281 [astro-ph.CO].
%
%
%
\bibitem{generic}
  K.~Hinterbichler, A.~Nicolis and M.~Porrati,
  JHEP {\bf 0909}, 089 (2009)
  [arXiv:0905.2359 [hep-th]].
  %
  E.~Babichev, C.~Deffayet and R.~Ziour,
  Int.\ J.\ Mod.\ Phys.\  D {\bf 18}, 2147 (2009)
  [arXiv:0905.2943 [hep-th]].
  %
  E.~Dyer and K.~Hinterbichler,
  JHEP {\bf 0911}, 059 (2009)
  [arXiv:0907.1691 [hep-th]].
  %
%
\bibitem{Endlich:2010zj}
  S.~Endlich, K.~Hinterbichler, L.~Hui, A.~Nicolis and J.~Wang,
  arXiv:1002.4873 [hep-th].
%
%
%
\bibitem{shandera}
  L.~Leblond and S.~Shandera,
  JCAP {\bf 0808}, 007 (2008)
  [arXiv:0802.2290 [hep-th]].
  C.~Cheung, P.~Creminelli, A.~L.~Fitzpatrick, J.~Kaplan and L.~Senatore,
  JHEP {\bf 0803}, 014 (2008)
  [arXiv:0709.0293 [hep-th]].
%
%
\bibitem{Nicolis:2004qq}
  A.~Nicolis and R.~Rattazzi,
  JHEP {\bf 0406}, 059 (2004)
  [arXiv:hep-th/0404159].
%
%
%
\bibitem{Vainshtein:1972sx}
  A.~I.~Vainshtein,
  Phys.\ Lett.\  B {\bf 39} (1972) 393.
%
%
\bibitem{Adams:2006sv}
  A.~Adams, N.~Arkani-Hamed, S.~Dubovsky, A.~Nicolis and R.~Rattazzi,
  JHEP {\bf 0610}, 014 (2006)
  [arXiv:hep-th/0602178].
  A.~Nicolis, R.~Rattazzi and E.~Trincherini,
  arXiv:0912.4258 [hep-th].
%
%
%
%
\bibitem{Lovelock:1971yv}
  D.~Lovelock,
  J.\ Math.\ Phys.\  {\bf 12}, 498 (1971).
%
%
\bibitem{Davis:2002gn}
  S.~C.~Davis,
  Phys.\ Rev.\  D {\bf 67}, 024030 (2003)
  [arXiv:hep-th/0208205].
%
\end{thebibliography}
\end{document}